\documentclass[a4paper]{JHEP3}
\usepackage[centertags]{amsmath}
\usepackage{amssymb}
\usepackage{graphicx}
\usepackage{amsthm}
\newtheorem*{thm*}{Theorem}

\title{Anomaly Induced Transport in Arbitrary Dimensions}

\author{\ R. Loganayagam\footnote{nayagam@physics.harvard.edu}\\
\small{\emph{Junior Fellow, Harvard Society of Fellows,}} \\
\small{\emph{Harvard University, Cambridge, MA 02138 .}} 
}

\abstract{
Motivated by  the consistency of a global anomaly with 
the second law of thermodynamics, we propose a form for
the anomaly induced charge/energy transport in arbitrary
even dimensions. In a given dimension, this form exhausts 
all second law constraints on anomaly induced transport 
at any given order in hydrodynamic derivative expansion.
We achieve by solving the second law constraints 
off-shell without resorting to hydrodynamic equations 
at lower orders. We also study various possible 
finite temperature corrections to such anomaly 
induced transport coefficients.
}

\preprint{}

\begin{document}

\section{Introduction}
One of the interesting features to come out of studying 
hydrodynamics of quantum field theories via holography
\cite{Kovtun:2004de,Bhattacharyya:2008jc}
is the discovery that there are macroscopic transport
coefficients in the hydrodynamics which probe the global
anomalies in the microscopic theory\cite{Bhattacharyya:2007vs,
Erdmenger:2008rm,Banerjee:2008th,Torabian:2009qk}. Since the anomalies
are exact objects in field theory, this leads to the 
hope that there are certain transport coefficients  of
interacting field theories which can be exactly computed
by relating them to underlying anomalies.

Son and Surowka in \cite{Son:2009tf} constrained various
such first order transport coefficients using the consistency of the
(3+1)d-anomaly with the second law of thermodynamics. Motivated
by the form of their answer, we propose a generalization
of the anomaly induced charge and energy transport in 
arbitrary dimensions and show that the form is consistent
with the second law\footnote{Recently, there have been indications that there 
might be additional transport \cite{Neiman:2010zi} induced
at finite temperature by the mixed gravitational 
anomalies\cite{Amado:2011zx,Landsteiner:2011cp}
in (3+1)dimensions at least in free theories. All these results 
generalize the results of various old calculations
in the literature\cite{Vilenkin:1978hb,Vilenkin:1979ui,
Vilenkin:1980fu,Vilenkin:1980zv,Vilenkin:1980ft,Vilenkin:1995um}.
However, a general argument linking these transport to 
mixed anomalies in an arbitrary interacting system is 
presently lacking (at least the author is unaware of any
such argument in the literature). 

In this article, we will neglect such additional 
contributions and focus only on purely global anomalies.
However it is possible that the succinct forms of the 
global anomaly induced transport presented in this paper
will lead to a suitable ansatz taking into account 
gravitational anomalies.}.

We will argue that the simple form of anomaly induced transport
presented in this paper exhausts all constraints on  anomaly induced transport
(at arbitrary order in hydrodynamic derivative expansion)
that can ever be derived from  second law considerations alone
(in a flat spacetime). This is achieved by solving the second law constraints 
off-shell without resorting to hydrodynamic equations 
at lower orders in derivative expansion\footnotemark. 

\footnotetext{This of course leaves open the possibility that there are additional 
transport terms which are in general induced by the anomaly
but are not captured by a second-law type argument.
Such terms, if present, would be beyond the scope of this
paper and their understanding would then crucially 
depend on finding a field theoretical way to derive these
transport coefficients in general interacting QFTs.}

The particular form of anomaly induced transport that we 
propose has various structural features and we resort to
a specific notation which exhibits these features clearly.
We begin by establishing the relevant notation/formalism in section
 \S\ref{sec:notation} . In particular, we introduce a useful 
notation which streamlines the entire discussion. 

We present the advertised generalization 
in the section \S\ref{sec:generald} and prove its consistency
with the second law. In the next section \S\ref{sec:son-surowka}
we restrict to (3+1)dimensions and show that under particular
frame redefinition, the results in \cite{Son:2009tf} 
are recovered. We conclude in \S\ref{sec:concl} with the 
discussion of various issues and future directions.

In the appendix \S\ref{app:ambiguity}, we discuss possible finite
temperature corrections to the anomaly-induced transport . These 
corrections represent other possible non-dissipative transport 
which are very similar in their structure to anomaly induced 
transport phenomena and are interesting in their own right.

In the next appendix \S\ref{app:explicit},
we present various expressions relevant for dimensions upto
$d=10$ for ready reference. The last appendix \S\ref{app:notation} 
is a summary of various notations employed in this paper.

\emph{Note:} As this paper was readied for submission, a preprint 
addressing similar questions \cite{Kharzeev:2011ds} appeared on 
arXiv.
 
\section{Basic formalism}\label{sec:notation}
We are interested in hydrodynamics of field theories in arbitrary even
spacetime dimensions. Let us denote the spacetime dimension by $2n$.
We will assume that the field theory has a global symmetry which has
a global anomaly - by this we mean, the covariant divergence of the 
covariant global current\footnote{In general, the currents obtained by
a direct variation of the path integral (the ''consistent'' currents)
are \emph{not} covariant under the shift of the non-dynamical gauge
fields. The covariant currents are  obtained by adding by hand an
additional current contribution called the Bardeen-Zumino current
\cite{Bardeen:1984pm}. Note that such ad-hoc additions are  allowed
in global symmetry currents and would have been illegitimate for
any current coupled to a dynamical gauge field.} is given by a series of terms involving the non-dynamical
gauge fields associated with this global symmetry. This in turn 
means that we can use the standard `covariant anomaly' in our 
discussion - with the anomaly symmetrically shared between the currents\footnotemark.

\footnotetext{Note that if we wanted to give dynamics to a subset of gauge fields
perturbatively, then (assuming we start from a path-integral measure which 
gives the symmetric `consistent anomaly') we have to add local 
counterterms (called Bardeen counterterms) made purely out of gauge fields 
in the microscopic action so that we shift the 
anomalies away from the gauged subgroup (A ready reference for Bardeen
counterterms in arbitrary dimensions is \cite{Chu:1996fr}). Any counter-term can be thought of
as a redefinition of the path-integral measure, and in this case we are just
redefining the measure so that it is invariant under the gauged subgroup. 

So, whenever a subgroup is gauged, all currents get corrected due to a 
Bardeen current coming from these additional terms which is a function of
various gauge fields. These Bardeen counterterms depend on the
details of the subgroup that is gauged and the effect of these terms
on the final hydrodynamics depends further on the details of the 
quantum gauge dynamics at finite temperature/chemical potential. 

The results derived here might be of some relevance to such  mixed anomalies
between global symmetries and gauge redundancies only if this gauge dynamics
is perturbative, i.e., if the theory is `weakly gauged' in which case the
effect of Bardeen terms can be accounted for perturbatively. }

We want to study hydrodynamics at finite temperature and cartan chemical
potentials for the global symmetry. We want to do this in the presence of the
non-dynamical gauge fields also turned on in the same cartan subgroup\footnote{
Note that these non-dynamical gauge fields violate the conservation of the global
charges. Thus, by introducing chemical potentials for these charges, we are 
making an implicit assumption that this non-conservation is in some sense `small'
enough that it makes sense to talk about local thermal/chemical equilibrium within 
each fluid element.}. When a global symmetry has an anomaly, the 
constitutive relations in the ideal hydrodynamics are modified - 
we will find it convenient to present the modifications in the following 
form for the energy, charge and the entropy currents
\footnote{Note that we do not include any tensor corrections to the energy
momentum tensor due to the anomalous transport. As will be seen later, the
second law considerations do not constrain such terms and hence such terms
do not fall into the ambit of the anomalous transport terms considered in this
paper. It is of course possible that more microscopic considerations lead
to additional transport phenomena which a second law argument is blind to. 
Further, in specific theories, anomalies might, in consonance with other
features of the theory, lead to specific phenomena which cannot be captured
by any such general analysis. All subsequent analysis is subject to these
caveats. The author wishes to thank Jyotirmoy Bhattacharya
for emphasizing this point.}  
\begin{equation}
\begin{split}
T^{\mu\nu} &\equiv \varepsilon u^\mu u^\nu + p P^{\mu\nu} + q^\mu_{anom}u^\nu + u^\mu q^\nu_{anom} + T^{\mu\nu}_{diss}\\
J^{i\mu} &\equiv n^i u^\mu + J^{i\mu}_{anom}+J^{i\mu}_{diss} \\
J^\mu_S &\equiv s u^\mu + J^\mu_{S,anom}+J^\mu_{S,diss}\\
\end{split}
\end{equation}
where $P^{\mu\nu}\equiv g^{\mu\nu}+u^\mu u^\nu$ , pressure of the fluid is $p$ and $\{\epsilon,n^i,s\}$ are the 
energy,charge and the entropy densities respectively. 

$u^\mu$ is the velocity of the fluid under consideration which obeys $u^\mu u_\mu =-1$.
We have denoted by $\{q^\mu_{anom},J^{i\mu}_{anom},
J^\mu_{S,anom}\}$ the anomalous heat/charge/entropy currents and by $\{T^{\mu\nu}_{diss},J^{i\mu}_{diss},
J^\mu_{S,diss}\}$ the dissipative currents. These are assumed to satisfy\footnote{
Note that we do not insist on particular frame conventions that need to be used for the
dissipative parts of the constitutive relation. So, most of what we say would work in
any frame \emph{provided the frame convention does not explicitly forbid the forms in
which  anomalous pieces appear in our equation}. For example, one of the commonly employed
frames is the Landau frame where terms like $q^{(\mu}_{anom}u^{\nu)}$ are explicitly forbidden.
To figure out the anomaly-induced transport in such frames, one needs to first shift
to a frame which allows such terms, add the transport in the new frame and then shift 
back. The author wishes to thank Jyotirmoy Bhattacharya and Sayantani Bhattacharyya
for discussions related to this point.}
\begin{equation}
\begin{split}
u_\mu q^\mu_{anom}&= u_\mu J^{i\mu}_{anom} = u_\mu J^\mu_{S,anom} = 0\\
D_\mu J^{i\mu}&=\mathfrak{A}^i\\
D_\mu T^{\mu\nu} &= J^i_\mu F_i^{\nu\mu}
\end{split}
\end{equation}
where $D$ is the covariant derivative including gauge fields and $\mathfrak{A}^i$ is the global anomaly.

Now the rate of entropy production is given by
\begin{equation}
\begin{split}
T D_\mu J^\mu_S &=T D_\mu J^\mu_S + \mu_i \left[ D_\mu J^{i\mu}-\mathfrak{A}^i \right] + u_\nu\left[D_\mu T^{\mu\nu}- J^i_\mu F_i^{\nu\mu}\right]\\
&=u^\mu \left[T D_\mu s + \mu_i D_\mu n^i -D_\mu \varepsilon\right] + D_\mu u^\mu \left[T s + \mu_i n^i - (\varepsilon+p)\right]\\
& + J^{i\mu}_{anom}E_{i\mu}-\mu_i\mathfrak{A}^i+ T D_\mu J^\mu_{S,anom} +\mu_i D_\mu J^{i\mu}_{anom}-(D_\mu+a_\mu) q^\mu_{anom}\\
& + T D_\mu J^\mu_{S,diss} + \mu_i D_\mu J^{i\mu}_{diss} + u_\nu\left[D_\mu T^{\mu\nu}_{diss}- J^i_{\mu,diss} F_i^{\nu\mu}\right]\\
&=J^{i\mu}_{anom}E_{i\mu}-\mu_i\mathfrak{A}^i + T D_\mu J^\mu_{S,anom} +\mu_i D_\mu J^{i\mu}_{anom}-(D_\mu+a_\mu) q^\mu_{anom}\\
& + T D_\mu J^\mu_{S,diss} + \mu_i D_\mu J^{i\mu}_{diss} + u_\nu\left[D_\mu T^{\mu\nu}_{diss}- J^i_{\mu,diss} F_i^{\nu\mu}\right]\\
\end{split}
\end{equation}
where $E_{i\mu}\equiv F_{i\mu\nu}u^\nu$ is the rest frame electric field and
$a_\mu\equiv (u.D)u_\mu$ is the acceleration field. 

The second law of thermodynamics is the assertion that the RHS of the above equation be positive. If we 
assume that the anomalous transport terms do not contribute to the entropy production then we get the
standard second law constraint on the dissipative part when there is no anomaly, viz., 
assuming 
\begin{equation}\label{eq:2ndLawAnom}
\begin{split}
(D_\mu+a_\mu) q^\mu_{anom}-J^{i\mu}_{anom}E_{i\mu} = T D_\mu J^\mu_{S,anom} + \mu_i \left[D_\mu J^{i\mu}_{anom}-\mathfrak{A}^i\right] 
\end{split}
\end{equation}
we get the second law constraint as
\begin{equation}
\begin{split}
&  T D_\mu J^\mu_{S,diss} + \mu_i D_\mu J^{i\mu}_{diss} + u_\nu\left[D_\mu T^{\mu\nu}_{diss}- J^i_{\mu,diss} F_i^{\nu\mu}\right]\geq 0 \\
\end{split}
\end{equation}
Note that if we can find even one solution that satisfies equation\eqref{eq:2ndLawAnom}, we can use that
to reduce the second-law to the usual non-anomalous case . 

To be useful this way the solution should be an `off-shell' solution i.e.,
we should be able to show that it is a solution \emph{without any further 
use of the conservation equations} which involve dissipative contributions
from various orders\footnote{In particular, this is not true of the solution presented by Son and Surowka in
\cite{Son:2009tf} since they use the ideal fluid equations of motion.}.
The main result of this paper is to demonstrate by constructing a 
solution that this can always be done in arbitrary dimensions\footnote{
Note that if we succeed in finding one such solution without any entropy production we 
would have also demonstrated by construction that there is a piece of transport
linking second law and anomalies which is non-dissipative. Hence, our 
assumption about the non-dissipative nature of anomalous transport would
be justified in hindsight}.

It is convenient to write the equation\eqref{eq:2ndLawAnom} in terms of forms. To this end, introduce the
hodge dual $2n-1$ forms $\{\bar{J}_{S,anom},\bar{q}_{anom},\bar{J}^{i}_{anom}\}$. We remind
the reader that in $2n$ dimensions, given any $2n-1$ form $\bar{V}$ hodge-dual to $V_\mu$
and a 1-form $A_\mu$, we have 
\begin{equation}
\begin{split}
D\bar{V}&=(D_\mu V^\mu)\ \text{Vol}_{2n}\\
A\wedge\bar{V}&=-\bar{V}\wedge A = A_\mu V^\mu \ \text{Vol}_{2n}\\
\end{split}
\end{equation}
So the equation\eqref{eq:2ndLawAnom} can be recast as
\begin{equation}\label{eq:2ndLawAnomF}
\begin{split}
T D\bar{J}_{S,anom} = D\bar{q}_{anom}+a\wedge\bar{q}_{anom}+\bar{J}^{i}_{anom}\wedge E_{i}-\mu_i \left[D \bar{J}^{i}_{anom}-\bar{\mathfrak{A}}^i\right] 
\end{split}
\end{equation}
where $\bar{\mathfrak{A}}^i\equiv\mathfrak{A}^i\ \text{Vol}_{2n}$ is the 2n-form Hodge dual to
the 0-form $\mathfrak{A}^i$.  

Now, we turn to a more detailed analysis of the anomaly. Let $F_i$ be the field-strength
2-form and we have already defined the electric 1-form via $E_{i\mu}\equiv F_{i\mu\nu} u^\nu $.
We can do an electric-magnetic decomposition
\begin{equation} F_{i\mu\nu} -\left[ u_{\mu}E_{i\nu}-E_{i\mu}u_{\nu}\right] \equiv B_{i\mu\nu} \end{equation}
or in the language of forms
\begin{equation} F_i = B_i + u\wedge E_i  \end{equation}
where $B_i$ is the magnetic 2-form completely transverse to $u^\mu$, i.e., $B_{i\mu\nu}u^\nu=0$.
In fact if our spacetime dimension is $2n$, in the rest frame of the fluid
${B}_i$ can be thought of as 2-form in $2n-1$ spatial dimensions. This in 
particular means that
\begin{equation} {B}_{i_1}\wedge {B}_{i_2} \wedge {B}_{i_3} \ldots\wedge {B}_{i_n} = 0 \end{equation}
where $i_k$s are the flavor indices.

We will also use the standard decomposition of the velocity gradients
\begin{equation} D_\mu u_\nu = \sigma_{\mu\nu}+\omega_{\mu\nu}-u_{\mu} a_\nu +\frac{\theta}{d-1}P_{\mu\nu} \end{equation}
in terms of the shear strain rate $\sigma_{\mu\nu}$, the vorticity $\omega_{\mu\nu}$,
the acceleration $a_\mu$ and the expansion rate $\theta$ of the fluid. This in 
particular means the exterior derivative of the velocity 1-form has
the decomposition
\begin{equation} Du = 2\omega -u\wedge a \end{equation}
where $\omega$ is the vorticity 2-form. Further, using $DF_i=D^2u=0$, we get
\begin{equation}\label{eq:DBDw}
\begin{split}
D{B}_i &= DF_i -Du\wedge E_i + u\wedge DE_i\\
&= -Du\wedge E_i + u\wedge DE_i \\
&= -2\omega\wedge E_i + u\wedge (DE_i+a\wedge E_i)\\
2D\omega&= Du\wedge a-u\wedge Da = 2\omega\wedge a-u\wedge Da\\
D(B_i+2\mu_i\omega) &= -2\omega\wedge(E_i-D\mu_i-a\mu_i)+u\wedge(DE_i+a\wedge E_i-Da) \\
\end{split}
\end{equation}

We can write the anomaly in $2n$ dimensions in the form
\begin{equation}
\begin{split}
\bar{\mathfrak{A}}^{i_0} &= \frac{1}{n!} \mathfrak{C}^{i_0 i_1 i_2 i_3 \ldots i_n } F_{i_1}\wedge F_{i_2} \wedge F_{i_3} \ldots\wedge F_{i_n}\\
& = \frac{1}{(n-1)!} \mathfrak{C}^{i_0 i_1 i_2 i_3 \ldots i_n }
  {B}_{i_1} \wedge{B}_{i_2} \wedge {B}_{i_3} \ldots\wedge {B}_{i_{n-1}}\wedge u\wedge E_{i_n}\\
\end{split}
\end{equation}
where $\mathfrak{C}^{i_0 i_1 i_2 i_3 \ldots i_n }$ is the anomaly coefficient which is completely 
symmetric in all its indices. 

All our subsequent expressions will be linear in the anomaly coefficient. We will
exploit this by introducing a notational trick which simplifies the presentation.
We will begin by doing a replacement 
\[ \mathfrak{C}^{i_0i_1\ldots i_n} \mapsto t^{i_0} t^{i_1}\ldots t^{i_n} \]
where we have replaced the anomaly coefficient with a bunch of fictitious parameters. This
turns the above equation to 
\begin{equation}
\begin{split}
\bar{\mathfrak{A}}^{i} &= t^{i}\frac{(t.F)^n}{n!}  = t^{i}\frac{(t.B)^{n-1}}{(n-1)!} \wedge u\wedge (t.E)\\
\end{split}
\end{equation}
we can always restore the original expression by doing the reverse replacement 
\[  t^{i_0} t^{i_1}\ldots t^{i_n} \mapsto \mathfrak{C}^{i_0i_1\ldots i_n} \]
We will further find it convenient to do a formal sum over all the even dimensions, writing\footnotemark
\begin{equation}
\begin{split}
\bar{\mathfrak{A}}^{i} &= t^{i}e^{t.F}  = t^{i}e^{t.B} \wedge u\wedge (t.E)\\
\end{split}
\end{equation}
Hence, in the following, to obtain the formulae for $2n$ dimensions we 
have to Taylor-expand in $t^i$s, take the terms with $n+1$ number 
of $t^i$s and do the replacement
\[  t^{i_0} t^{i_1}\ldots t^{i_n} \mapsto \mathfrak{C}^{i_0i_1\ldots i_n} \]
where $\mathfrak{C}^{i_0i_1\ldots i_n}$is the anomaly-coefficient of the theory.
\footnotetext{ Note that we are denoting the formal sum over arbitrary dimensions by the same
symbols as those used to denote quantities in a specific dimension. This is analogous 
to the notation used when say we want to represent anomaly polynomials by formal sum
over all even dimensions. We hope the reader does not find this too confusing - 
our main motivation for this notation is  that we believe it  more clearly 
exhibits various structures associated with the anomaly-induced transport.} 

\section{Anomalous transport in General dimensions}\label{sec:generald}
We now want to present a solution for the equation\eqref{eq:2ndLawAnomF} which we repeat
here
\begin{equation}
\begin{split}
T D\bar{J}_{S,anom}+\mu_i \left[D \bar{J}^{i}_{anom}-\bar{\mathfrak{A}}^i\right]  = D\bar{q}_{anom}+a\wedge\bar{q}_{anom}+\bar{J}^{i}_{anom}\wedge E_{i}
\end{split}
\end{equation}
We remind the reader that this is physically the statement that the anomalous
transport leads to no entropy production. One simple solution of the above 
equation valid in arbitrary dimensions is\footnote{Note that while these solutions
as presented have $\omega$ in their denominators, they have a smooth $\omega\to 0$ limit.}
\begin{equation}\label{eq:solnD}
\begin{split}
\bar{q}_{anom} &= -\left[\frac{(2t.\mu\ \omega-1)e^{t.(B+2\mu \omega)}+e^{t.B}}{4\omega^2} \right]\wedge u \\
\bar{J}^i_{anom} &=- t^i\left[\frac{e^{t.(B+2\mu \omega)}-e^{t.B}}{2\omega} \right]\wedge u\\
\bar{J}_{S,anom} &=0
\end{split}
\end{equation}
where we have used the notation introduced in the previous section to present the solution in 
arbitrary dimensions in a nice succinct form. This is the central result of this paper
and we will now demonstrate that this is indeed a solution. In our notation, 
this is straightforward . Using the results in equation\eqref{eq:DBDw}, we have
\begin{equation}\label{eq:DqDJ}
\begin{split}
D&\bar{q}_{anom}+a\wedge\bar{q}_{anom}\\
&= \left\{\frac{\left[(2t.\mu\ \omega-1)t.(E-D\mu-a\mu)-(D\mu+a\mu)\right]e^{t.(B+2\mu \omega)}+(t.E)e^{t.B}}{2\omega} \right\}\wedge u \\
&\qquad -\left[\frac{(2t.\mu\ \omega-1)e^{t.(B+2\mu \omega)}+e^{t.B}}{2\omega} \right]\\
\end{split}
\end{equation}
\begin{equation}\label{eq:DqDJ2}
\begin{split}
D&\bar{J}^i_{anom}\\
&= t^i \left[t.(E-D\mu-a\mu)\ e^{t.(B+2\mu \omega)}-(t.E)e^{t.B}\right]\wedge u\\
&\qquad -t^i\left[e^{t.(B+2\mu \omega)}-e^{t.B}\right]\\
\end{split}
\end{equation}
\begin{equation}\label{eq:DqDJ3}
\begin{split}
\mu_i&\left[D\bar{J}^i_{anom}-t^i e^{t.B}\wedge u\wedge(t.E) \right]-\left[ D\bar{q}_{anom}+a\wedge\bar{q}_{anom}+ \bar{J}^i_{anom}\wedge E_i\right]\\
&= -\left[\frac{e^{t.(B+2\mu \omega)}-(2t.\mu\ \omega+1)e^{t.B}}{2\omega} \right]
\end{split}
\end{equation}
Note that in the last line every term is a $2n$ form in $2n$ dimensions made only out of 
the `spatial' forms $\omega$ and $B_i$ and any such form vanishes. Hence, we conclude
that this is indeed the required `offshell' solution.
 
Further it is interesting to note that our solution satisfies a First law-type  relation
\begin{equation}
\begin{split}
\frac{\partial\bar{q}_{anom}}{\partial \mu_i}= \mu_k \frac{\partial\bar{J}^k_{anom}}{\partial \mu_i} =- t^i(t.\mu)\left[e^{t.(B+2\mu \omega)}\right]\wedge u \\
\end{split}
\end{equation}
and a generalized Onsager- type Reciprocity relation 
\begin{equation}
\begin{split}
\frac{\delta\bar{q}_{anom}}{\delta B_i}= \frac{1}{2}\frac{\delta\bar{J}^i_{anom}}{\delta \omega}=-t^i\left[\frac{(2t.\mu\ \omega-1)e^{t.(B+2\mu \omega)}+e^{t.B}}{4\omega^2} \right]\wedge u  \\
\end{split}
\end{equation}
This means we can think of both the heat and the charge current as generated from a
formal object $\mathcal{V}_{anom}$
\begin{equation}
\begin{split}
\bar{q}_{anom} &=\frac{1}{2}\frac{\delta \mathcal{V}_{anom}}{\delta \omega}\wedge u\\
\bar{J}^i_{anom} &= \frac{\delta\mathcal{V}_{anom}}{\delta B_i}\wedge u  \\
\mathcal{V}_{anom} &\equiv -\left[\frac{e^{t.(B+2\mu \omega)}-e^{t.B}}{2\omega}\right] \\
\end{split}
\end{equation}
Note that $\mathcal{V}_{anom}$ is a formal spatial $2n$-form in $2n$ dimensions whose role is similar to
the anomaly polynomial which is a formal $2n+2$ form in $2n$ dimensions.

Further, we can think of these currents as derived from a single expression for the 
anomaly induced Gibbs free energy current $\bar{\mathcal{G}}_{anom}$
\begin{equation}
\begin{split}
\bar{\mathcal{G}}_{anom} &= \left[\frac{e^{t.(B+2\mu \omega)}-e^{t.B}(2t.\mu\ \omega+1)}{(2\omega)^2}\right]=-\frac{1}{2\omega} \left[\mathcal{V}_{anom}-\mathcal{V}_{anom}(\omega=0)\right]\wedge u \\
\bar{J}^i_{anom} &= -\frac{\partial\bar{\mathcal{G}}_{anom}}{\partial \mu_i}  \\
\bar{J}_{S,anom} &=-\frac{\partial\bar{\mathcal{G}}_{anom}}{\partial T}\\
\bar{q}_{anom} &=\bar{\mathcal{G}}_{anom}+T\bar{J}_{S,anom}+\mu_i\bar{J}^i_{anom}\\
\end{split}
\end{equation}

\section{Comparison with Son-Surowka result}\label{sec:son-surowka}
In $d=4$ case (when $n=2$) we get the following result for the anomalous transport
\begin{equation}\label{eq:soln4Alt}
\begin{split}
\bar{\mathfrak{A}}^i &= \frac{1}{2!}\mathfrak{C}^{ijk}F_j\wedge F_k \\
\bar{q}_{anom} &= -\mathfrak{C}^{ijk}\mu_i\mu_j\left[\frac{1}{2} B_k + \frac{2}{3}\mu_k\omega\right]\wedge u \\
\bar{J}^i_{anom} &=-\mathfrak{C}^{ijk}\mu_j\left[ B_k + \mu_k\omega\right]\wedge u \\
\bar{J}_{S,anom} &=0
\end{split}
\end{equation}
we take the Hodge-duals to get
\begin{equation}\label{eq:soln4Hodge}
\begin{split}
\mathfrak{A}^i &= \frac{1}{(2!)^3}\mathfrak{C}^{ijk}\epsilon^{\mu\nu\lambda\sigma}F_{j\mu\nu}\wedge F_{k\lambda\sigma} \\
{q}^\mu_{anom} &= -\frac{1}{2!}\mathfrak{C}^{ijk}\mu_i\mu_j\epsilon^{\mu\nu\lambda\sigma}\left[\frac{1}{2} B_k + \frac{2}{3}\mu_k\omega\right]_{\nu\lambda}u_\sigma \\
{J}^{i\mu}_{anom} &=-\frac{1}{2!}\mathfrak{C}^{ijk}\mu_j\epsilon^{\mu\nu\lambda\sigma}\left[ B_k + \mu_k\omega\right]_{\nu\lambda}u_\sigma\\
{J}^\mu_{S,anom} &=0
\end{split}
\end{equation}
To compare with \cite{Son:2009tf}, we first define 
\[ \mathfrak{C}^{ijk}\equiv-{C}^{ijk},\ \quad \bar{B}_i^\mu \equiv \frac{1}{2!} \epsilon^{\mu\nu\lambda\sigma} B_{i\nu\lambda}u_\sigma \quad\text{and}\quad \bar{\omega}^\mu \equiv \frac{1}{2!} \epsilon^{\mu\nu\lambda\sigma} \omega_{\nu\lambda}u_\sigma \]
to get
\begin{equation}\label{eq:soln4Son}
\begin{split}
\mathfrak{A}^i &= -\frac{1}{8}{C}^{ijk}\epsilon^{\mu\nu\lambda\sigma}F_{j\mu\nu}\wedge F_{k\lambda\sigma} \\
{q}^\mu_{anom} &= {C}^{ijk}\mu_i\mu_j\left[\frac{1}{2} \bar{B}_k^\mu + \frac{2}{3}\mu_k\bar{\omega}^\mu\right] \\
{J}^{i\mu}_{anom} &={C}^{ijk}\mu_j\left[ \bar{B}_k^\mu + \mu_k\bar{\omega}^\mu\right]\\
{J}^\mu_{S,anom} &=0
\end{split}
\end{equation}
Son and Surowka presented their results in the Landau frame (where $q^\mu=0$) to first order in derivative expansion.
The most general frame-change at this order leads to
\begin{equation}\label{eq:frameChange}
\begin{split}
u^\mu&\mapsto u^\mu+\delta u^\mu+\ldots \\
q^\mu&\mapsto q^\mu + (\varepsilon+p)\delta u^\mu +\ldots \\
J^{i\mu} &\mapsto J^{i\mu} + n^i \delta u^\mu +\ldots \\
J^{\mu}_S &\mapsto J^{\mu}_S + s \delta u^\mu +\ldots \\
\end{split}  
\end{equation}
Hence to set  $q^\mu=0$ we choose $\delta u^\mu = -\frac{q^\mu}{\varepsilon+p}$ which gives
\begin{equation}\label{eq:soln4Landau}
\begin{split}
\mathfrak{A}^i &= -\frac{1}{8}{C}^{ijk}\epsilon^{\mu\nu\lambda\sigma}F_{j\mu\nu}\wedge F_{k\lambda\sigma} \\
{q}^\mu_{anom} &= 0 \\
{J}^{i\mu}_{anom} &={C}^{ijk}\mu_j\left[ \bar{B}_k^\mu + \mu_k\bar{\omega}^\mu\right]-\frac{n^i}{\varepsilon+p}{C}^{ljk}\mu_l\mu_j\left[\frac{1}{2} \bar{B}_k^\mu + \frac{2}{3}\mu_k\bar{\omega}^\mu\right]+\ldots\\
{J}^\mu_{S,anom} &=-\frac{s}{\varepsilon+p}{C}^{ijk}\mu_i\mu_j\left[\frac{1}{2} \bar{B}_k^\mu + \frac{2}{3}\mu_k\bar{\omega}^\mu\right]+\ldots\\
&=-\frac{1}{T}{C}^{ijk}\mu_i\mu_j\left[\frac{1}{2} \bar{B}_k^\mu + \frac{2}{3}\mu_k\bar{\omega}^\mu\right]\\
&\qquad+\frac{\mu_i n^i}{T(\varepsilon+p)}{C}^{ijk}\mu_i\mu_j\left[\frac{1}{2} \bar{B}_k^\mu + \frac{2}{3}\mu_k\bar{\omega}^\mu\right]+\ldots\\
&=-\frac{1}{T}{C}^{ijk}\mu_i\mu_j\left[\frac{1}{2} \bar{B}_k^\mu + \frac{2}{3}\mu_k\bar{\omega}^\mu\right]\\
&\qquad-\frac{\mu_i{J}^{i\mu}_{anom}}{T}+\frac{1}{T}{C}^{ijk}\mu_i\mu_j\left[ \bar{B}_k^\mu + \mu_k\bar{\omega}^\mu\right]+\ldots\\
&=\frac{1}{T}{C}^{ijk}\mu_i\mu_j\left[\frac{1}{2} \bar{B}_k^\mu + \frac{1}{3}\mu_k\bar{\omega}^\mu\right]-\frac{\mu_i{J}^{i\mu}_{anom}}{T}+\ldots\\
\end{split}
\end{equation}
which is exactly the expression obtained in \cite{Son:2009tf}.

Certain comments are in order - as the manipulations which lead to the above
expressions make it clear, the expressions above get corrected at the next 
order in derivative expansion and in all subsequent orders. To find these
corrections in the Landau frame, one needs to know the higher derivative pieces in the 
constitutive relation and th resultant corrections to the equations of motion. 
- hence, this is an `on-shell' solution unlike the `off-shell' solution we started with
\footnotemark.The beauty of the solution presented in our frame 
is that the solution is independent of all such detailed transport 
coefficients at lower orders in derivative expansion.
\footnotetext{
This statement holds true even for the additional terms proportional to temperature 
 proposed in \cite{Neiman:2010zi}.
See appendix \ref{app:ambiguity} for a more detailed discussion of such extra terms in arbitrary dimensions.
}

This exercise can be repeated in arbitrary dimensions in the Landau frame and the conclusions
are similar. Whereas the leading order pieces in the anomalous transport can be determined
via leading order equations of motion, the subleading orders require the knowledge of more
pieces in the constitutive relation. This difficulty is again an artifact of working
in the Landau frame : our solution gives a way of relaxing the second law constraints
back to the non-anomalous case in most other frames. Given the complete constitutive
relation, one can always frame-transform our solution to the Landau frame to get the
anomalous transport to any order one wants.

\section{Discussion}\label{sec:concl}

In this paper, we have proposed  a form for the anomaly-induced transport in arbitrary
even dimensions motivated by consistency with the second law. This transport takes a 
very suggestive form and it is tempting to speculate that this is related to
some sort of an index-like object, especially since the anomalies themselves
are computed by various indices. 

We will present a particular suggestion for what this index might be.  In presence of anomaly,
the rotational equilibrium states get deformed because of the anomaly-induced transport. 
By taking a thermal density matrix of free chiral fermions  on  $S^{2n-1}\times R$ with 
chemical potentials for angular momentum/charges turned on, we should be able to 
see the transport proposed in this paper\footnote{Note that this is
a very similar  computation to the ones performed by Vilenkin \cite{Vilenkin:1978hb,Vilenkin:1979ui,
Vilenkin:1980fu,Vilenkin:1980zv,Vilenkin:1980ft,Vilenkin:1995um} in flat-space in (3+1)d.}.
This suggests that we are looking for a related index of some bundle on $S^{2n-1}\times S^1$ 
with various appropriate twists. It will be nice if we could make this proposal precise.
By fluid-gravity correspondence, similar rotational equilibrium states in large N
strongly coupled theories are dual to  blackhole solutions of
Einstein-Maxwell-Chern-Simons actions in global AdS$_{2n+1}$.
It is an interesting question whether these gravity solutions can be constructed
exactly\footnote{The author wishes to thank Mukund Rangamani for related discussions.} .

As we have mentioned at various points in the paper, we have completely ignored the 
issue of pure and mixed gravitational anomalies in our discussion. Hopefully, the simple
solution given in this paper and the methods used to construct it can be generalized to
gravitational anomalies. The structure of various finite temperature corrections (discussed
in Appendix \S\ref{app:ambiguity} ) and the $3+1$d free-theory computation of 
\cite{Landsteiner:2011cp} suggest relations between gravitational anomalies and
various finite temperature corrections , but there is no definitive argument (apart
from the statement that they have similar symmetry structure) to link these two.
It would be instructive to see whether for free theories in arbitrary dimensions
we still have a relation between finite temperature corrections to 
anomaly-induced transport on one hand and the coefficients of anomalies 
involving gravity on the other hand\footnote{This for example might be achievable
by computing these coefficients in kinetic theory of chiral fermion gases in 
arbitrary dimensions \cite{futurePiotr}.}. It is also worthwhile to 
independently understand these finite temperature
corrections in various theories, as some of them violate CP and hence might 
of some phenomenological relevance.

The discovery of anomalies and their various effects have led us to a deeper
understanding of QFTs with Lorentz-invariant ground states. While various
effects of anomalies in finite temperature and finite density have been explored
by now, we still lack a complete framework to encompass various anomaly-induced 
finite temperature/density transport. We hope that this work is a small step 
towards that broader goal.

\subsection*{Acknowledgements}
It is a pleasure to thank Jyotirmoy Bhattacharya, Sayantani Bhattacharyya,  
Sean Hartnoll, Nabil Iqbal, Tongyan Lin, John Mason, Vladimir Manucharyan,
Shiraz Minwalla, Mukund Rangamani, David Simmons-Duffin, Dam Thanh Son and
Piotr Surowka  for various useful discussions on ideas presented in this 
paper. I would like to thank ICTS, TIFR for their hospitality during 
the workshop on string theory and its applications organised at TIFR Mumbai. 
This work was supported by the Harvard Society of Fellows through a junior
fellowship. Finally, I would  like to thank various colleagues at the 
society for interesting discussions.

\section*{Appendices}
\appendix

\section{Finite temperature corrections to anomaly-induced transport}\label{app:ambiguity}
In the main text of this article we presented a solution to the equation 
\eqref{eq:2ndLawAnomF} 
\begin{equation}\label{eq:Anom2App}
\begin{split}
T D\bar{J}_{S,anom}+\mu_i \left[D \bar{J}^{i}_{anom}-\bar{\mathfrak{A}}^i\right]  = D\bar{q}_{anom}+a\wedge\bar{q}_{anom}+\bar{J}^{i}_{anom}\wedge E_{i}
\end{split}
\end{equation}
with $\bar{\mathfrak{A}}^i\equiv t^i e^{t.B}\wedge u\wedge (t.E)$ in the form
\begin{equation}
\begin{split}
\bar{q}_{anom} &= -\left[\frac{(2t.\mu\ \omega-1)e^{t.(B+2\mu \omega)}+e^{t.B}}{4\omega^2} \right]\wedge u \\
\bar{J}^i_{anom} &=- t^i\left[\frac{e^{t.(B+2\mu \omega)}-e^{t.B}}{2\omega} \right]\wedge u\\
\bar{J}_{S,anom} &=0
\end{split}
\end{equation}

In this appendix, we will examine the uniqueness of this solution. Since we are solving an
inhomogeneous linear equation, the difference between various solutions of this equation 
satisfies the \emph{homogeneous} equation
\begin{equation}
\begin{split}
T D\bar{J}_{S,rev}+\mu_i D \bar{J}^{i}_{rev} = D\bar{q}_{rev}+a\wedge\bar{q}_{rev}+\bar{J}^{i}_{rev}\wedge E_{i}
\end{split}
\end{equation}
We have used the suffix $rev$ to denote that physically these denote transport processes which do not
result in entropy production and hence are reversible. Evidently, it is too ambitious to try to classify
\emph{all} possible reversible transport processes which can happen in hydrodynamics, so we will
content ourselves with describing some classes of solutions to the above equations which are 
close in form to the transport due to global anomalies.

In examining the solution-space of the equation~\eqref{eq:Anom2App}, it is insightful to first prove
the following theorem
\begin{thm*}
Consider a $\mathcal{V}_{anom}$ of the form
\begin{equation} \mathcal{V}_{anom} \equiv -\frac{(\mathfrak{f}[B_i+2\mu_i\omega]-\mathfrak{f}[B_i])}{2\omega} -T^2 \omega\wedge \mathfrak{g}[B_i+2\mu_i\omega, T\omega] \end{equation}
where $\mathfrak{f}$ and $\mathfrak{g}$ are some Taylor-expandable functions\footnotemark
at the origin (i.e., are analytic at the origin). Construct from this the following currents
\begin{equation}\label{eq:Jrevs}
\begin{split}
\bar{\mathcal{G}}_{rev} &=-\frac{1}{2\omega} \left[\mathcal{V}_{anom}-\mathcal{V}_{anom}(\omega=0)\right]\wedge u\\
\bar{q}_{rev} &=\frac{1}{2}\frac{\delta \mathcal{V}_{anom}}{\delta \omega}\wedge u=\bar{\mathcal{G}}_{rev}+T\bar{J}_{S,rev}+\mu_i\bar{J}^i_{rev}\\
\bar{J}^i_{rev} &= \frac{\delta\mathcal{V}_{anom}}{\delta B_i}\wedge u=-\frac{\partial\bar{\mathcal{G}}_{rev}}{\partial \mu_i}   \\
\bar{J}_{S,rev} &= -\frac{\partial\bar{\mathcal{G}}_{rev}}{\partial T}\\
\end{split}
\end{equation}
Then they satisfy
\begin{equation}\label{eq:fundamental}
\begin{split}
T D\bar{J}_{S,rev}+\mu_i \left( D \bar{J}^{i}_{rev} -\frac{\delta^2 \mathfrak{f}[B_k]}{\delta B_i \delta B_j}\wedge u\wedge E_j\right) = D\bar{q}_{rev}+a\wedge\bar{q}_{rev}+\bar{J}^{i}_{rev}\wedge E_{i}
\end{split}
\end{equation}
 \footnotetext{If we only want
the parts relevant to $2n$ spacetime dimensions, we can take treat $\mathfrak{f}$ and 
$\mathfrak{g}$ as $(n+1)$th and $(n-1)$th degree homogeneous polynomials of
their arguments respectively. This makes $\mathcal{V}_{anom}$ into a formal spatial $2n$ form. }  
\end{thm*}
Notice that the main solution of this paper appears as a special case of this general theorem
(with $\mathfrak{f}[B_i]=e^{t.B}$ and $\mathfrak{g}=0$). At least in this subspace of solutions
this is the unique solution at zero temperature.

It can be checked by explicit computation that these class of  solutions satisfy first-law type relations
\begin{equation}
\begin{split}
\frac{\partial\bar{q}_{anom}}{\partial T}&= T \frac{\partial\bar{J}_{S,rev}}{\partial T}+\mu_k \frac{\partial\bar{J}^k_{rev}}{\partial T}  \\
\frac{\partial\bar{q}_{anom}}{\partial \mu_i}&= T \frac{\partial\bar{J}_{S,rev}}{\partial \mu_i}+\mu_k \frac{\partial\bar{J}^k_{rev}}{\partial \mu_i}  \\
\end{split}
\end{equation}
and a generalized Onsager- type Reciprocity relation 
\begin{equation}
\begin{split}
\frac{\delta\bar{q}_{rev}}{\delta B_i}= \frac{1}{2}\frac{\delta\bar{J}^i_{rev}}{\delta \omega} \\
\end{split}
\end{equation}

Now, it is clear that if we take $\mathfrak{f}=0$, we get a large class of homogeneous solutions that
we want. In $2n$ dimensions, $\mathfrak{g}$ takes the general form
\begin{equation}\label{eq:gFormula}
\begin{split}
\mathfrak{g} &= \sum_{k=0}^{n-1} \frac{(T\omega)^{n-1-k}}{k!(n-1-k)!}\wedge\mathfrak{g}_k^{i_1\ldots i_k} (B+2\mu\omega)_{i_1}\wedge\ldots(B+2\mu\omega)_{i_k}
\end{split}
\end{equation}
where $\mathfrak{g}_k^{i_1i_2\ldots i_k}$ is some invariant tensor of the global symmetry\footnote{
Curiously, such invariant tensors inevitably occur for odd $n-k$ (only odd $n-k$ occur if $\mathfrak{g}$
is taken to be an even function of $T\omega$ ) whenever there is a mixed 
global-gravitational anomaly of the form
\begin{equation}
\begin{split}
D\bar{J}^{i_1} \supset \sum_{k=1}^{n}\frac{\alpha_k}{k!(n-k+1)!}Tr(R^{n-k+1})\wedge  \mathfrak{g}_{k}^{i_1\ldots i_k}F_{i_2}\wedge\ldots\wedge F_{i_k} \
\end{split}
\end{equation}
hence it is interesting to speculate that theories with mixed anomalies 
generically have this kind of transport - in fact, something like this 
happens for free fermions in (3+1)dimensions \cite{Landsteiner:2011cp}
but the question of  whether such a relation continues to hold for 
strongly  coupled theories with same coefficient is open. 

More generally, there is the question of finding the transport consistent
with the second law in the presence of mixed anomalies which is yet to 
be answered.}. 

Note that this construction which we just outlined is a
generalization of the finite temperature corrections in (3+1)d 
proposed by the authors of \cite{Neiman:2010zi}. In (3+1)d, for example, $n=2$,
we can take 
\begin{equation}
\begin{split}
\mathfrak{g} &= \mathfrak{g}_0(T\omega)+ \mathfrak{g}_1^i(B+2\mu\omega)_i 
\end{split}
\end{equation}
which gives
\begin{equation}
\begin{split}
\bar{\mathcal{G}}_{rev}&=\left[\frac{1}{2}T^2\mathfrak{g}_1^i B_i +\left(\frac{1}{2}T^3\mathfrak{g}_0+T^2\mu_i\mathfrak{g}_1^i\right) \omega\right]  \wedge u\\
\bar{q}_{rev} &= -T^2\left[\frac{1}{2}\mathfrak{g}_1^i B_i + \left(T\mathfrak{g}_0+2\mathfrak{g}_1^i\mu_i\right)\ \omega \right]\wedge u \\
\bar{J}^i_{rev} &=- T^2\mathfrak{g}_1^i \omega \wedge u\\
\bar{J}_{S,rev} &= -T\left[\mathfrak{g}_1^i B_i + \left(\frac{3}{2}T\mathfrak{g}_0+2\mathfrak{g}_1^i\mu_i\right)\ \omega \right]\wedge u\\
\end{split}
\end{equation}
and these exactly correspond to the finite temperature corrections proposed in \cite{Neiman:2010zi},
provided we shift to the Landau frame using the procedure described in 
section \S\ref{sec:son-surowka}.

The (3+1)d free fermion calculation\footnotemark in \cite{Landsteiner:2011cp} gives $\mathfrak{g}_1^i\propto
tr\ T^i$ , i.e., $\mathfrak{g}_1^i$ is non-zero in free fermion theories if and only if 
there is a mixed gravitational anomaly. It is unknown what happens in interacting theories.
\footnotetext{Also see old calculations by Vilenkin in the context of neutrinos 
 \cite{Vilenkin:1978hb,Vilenkin:1979ui,
Vilenkin:1980fu,Vilenkin:1980zv,Vilenkin:1980ft,Vilenkin:1995um}.}
As far as the author is aware $\mathfrak{g}_0$ has never been calculated in free theory
or otherwise. Given that the associated transport violates CP, it might be of some
phenomenological relevance as a possible measure of CP violation (say in the early universe).  

A similar exercise can be repeated in arbitrary dimensions to get the finite temperature
corrections from the equations \eqref{eq:Jrevs} and \eqref{eq:gFormula} .

\section{Explicit expressions in various cases}\label{app:explicit}
In this appendix, we present the explicit solutions for $B=0$ and $\omega=0$ followed
by the explicit solutions for all even dimensions till d=10. 

All these follow from our solution - keeping only parts relevant to $d=2n$
dimensions
\begin{equation}\label{eq:soln2n}
\begin{split}
\bar{\mathfrak{A}}^i &= t^i\frac{(t.F)^{n}}{n!}\\
\mathcal{V}_{anom} &= -\frac{\left[t.(B+2\mu \omega)\right]^{n+1}-[t.B]^{n+1}}{(2\omega) (n+1)!}\\
\bar{\mathcal{G}}_{anom} &= \frac{\left[t.(B+2\mu \omega)\right]^{n+1}-[t.B]^{n+1}-(n+1)[2t.\mu \omega][t.B]^{n}}{(2\omega)^2 (n+1)!}\\
\bar{q}_{anom} &=  -\left[\frac{(n+1)[2t.\mu \omega] \left[t.(B+2\mu \omega)\right]^{n} - \left[t.(B+2\mu \omega)\right]^{n+1}+[t.B]^{n+1}}{(2\omega)^2 (n+1)!} \right]\wedge u \\
\bar{J}^i_{anom} &=- t^i\left[\frac{\left[t.(B+2\mu \omega)\right]^n-[t.B]^n}{(2\omega) n!} \right]\wedge u \\
\bar{J}_{S,anom} &=0
\end{split}
\end{equation}
\subsection*{B=0 case : Chiral Vortical Effect}
We now present the answer in arbitrary dimensions when the gauge fields are turned off
\begin{equation}\label{eq:solnBeq0}
\begin{split}
\bar{\mathfrak{A}}^i &= 0\\
\bar{\mathcal{G}}_{anom} &= \frac{1}{(n+1)!}\mathfrak{C}^{i_0i_1i_2\ldots i_n}\mu_{i_0}\mu_{i_1}\mu_{i_2}\ldots\mu_{i_n}(2\omega)^{n-1}\wedge u \\
\bar{q}_{anom} &= -\frac{n}{(n+1)!}\mathfrak{C}^{i_0i_1i_2\ldots i_n}\mu_{i_0}\mu_{i_1}\mu_{i_2}\ldots\mu_{i_n}(2\omega)^{n-1}\wedge u \\
\bar{J}^{i_0}_{anom} &= -\frac{1}{n!}\mathfrak{C}^{i_0i_1i_2\ldots i_n}\mu_{i_1}\mu_{i_2}\ldots\mu_{i_n}(2\omega)^{n-1}\wedge u \\
\bar{J}_{S,anom} &=0
\end{split}
\end{equation}
This gives the chiral vortical effect in arbitrary dimensions due to the anomaly. Note that it 
is still present when the anomaly itself has been turned off. 

\subsection*{$\omega$=0 case : Chiral Magnetic effect}
We now present the answer in arbitrary dimensions when the vorticity is turned off
\begin{equation}\label{eq:solnOmegaeq0}
\begin{split}
\bar{\mathfrak{A}}^i &= t^i\frac{(t.F)^{n}}{n!}\\
\bar{\mathcal{G}}_{anom} &=  \frac{1}{2!(n-1)!}\mathfrak{C}^{i_0i_1i_2\ldots i_n}\mu_{i_0}\mu_{i_1}B_{i_2}B_{i_3}\ldots B_{i_n}\wedge u \\
\bar{q}_{anom} &=  -\frac{1}{2!(n-1)!}\mathfrak{C}^{i_0i_1i_2\ldots i_n}\mu_{i_0}\mu_{i_1}B_{i_2}B_{i_3}\ldots B_{i_n}\wedge u \\
\bar{J}^i_{anom} &= -\frac{1}{(n-1)!}\mathfrak{C}^{i_0i_1i_2\ldots i_n}\mu_{i_1}B_{i_2}B_{i_3}\ldots B_{i_n}\wedge u \\
\bar{J}_{S,anom} &=0
\end{split}
\end{equation}
This gives the chiral magnetic effect in arbitrary dimensions due to the anomaly.

We now present the explicit solutions for all even dimensions till
d=10 for ready reference.

\subsection*{d=2}
\begin{equation}\label{eq:soln2}
\begin{split}
\bar{\mathfrak{A}}^i &= \mathfrak{C}^{ij}F_j \\
\bar{\mathcal{G}}_{anom} &=\frac{1}{2!}\mathfrak{C}^{jk}\mu^2_{jk} (2\omega)\wedge u \\
\mathcal{V}_{anom} &=-\mathfrak{C}^{jk}\left[\mu_j B_k + \mu^2_{jk} \omega\right] \\
\bar{q}_{anom} &=\frac{1}{2}\frac{\delta \mathcal{V}_{anom}}{\delta \omega}\wedge u= -\frac{1}{2}\mathfrak{C}^{ij}\mu^2_{ij} u \\
\bar{J}^i_{anom} &= \frac{\delta\mathcal{V}_{anom}}{\delta B_i}\wedge u =- \mathfrak{C}^{ij}\mu_j u \\
\bar{J}_{S,anom} &= \frac{1}{2\omega}\frac{\partial \mathcal{V}_{anom}}{\partial T}\wedge u=0
\end{split}
\end{equation}
\subsection*{d=4}
\begin{equation}\label{eq:soln4}
\begin{split}
\bar{\mathfrak{A}}^i &= \frac{1}{2!}\mathfrak{C}^{ijk}F^2_{jk} \\
\bar{\mathcal{G}}_{anom} &=\left[\frac{1}{3!}\mathfrak{C}^{ijk}\mu^3_{ijk} (2\omega)+ \frac{1}{2!}\mathfrak{C}^{ijk}\mu^3_{ij} B_k\right]\wedge u \\
\mathcal{V}_{anom} &=-\mathfrak{C}^{jkl}\mu_j\left[\frac{1}{2} B^2_{kl}+  B_k\wedge\mu_l\omega+\frac{2}{3}\mu^2_{kl}\omega^2 \right]\wedge u\\
\bar{q}_{anom} &=\frac{1}{2}\frac{\delta \mathcal{V}_{anom}}{\delta \omega}\wedge u= -\mathfrak{C}^{ijk}\mu^2_{ij}\left[\frac{1}{2} B_k + \frac{2}{3}\mu_k\omega\right]\wedge u \\
\bar{J}^i_{anom} &= \frac{\delta\mathcal{V}_{anom}}{\delta B_i}\wedge u =-\mathfrak{C}^{ijk}\mu_j\left[ B_k + \mu_k\omega\right]\wedge u \\
\bar{J}_{S,anom} &= \frac{1}{2\omega}\frac{\partial \mathcal{V}_{anom}}{\partial T}\wedge u=0
\end{split}
\end{equation}
\subsection*{d=6}
\begin{equation}\label{eq:soln6}
\begin{split}
\bar{\mathfrak{A}}^i &= \frac{1}{3!}\mathfrak{C}^{ijkl}F^3_{jkl}\\
\bar{\mathcal{G}}_{anom} &=\left[\frac{1}{4!}\mathfrak{C}^{ijkl}\mu^4_{ijkl} (2\omega)^2+\frac{1}{3!}\mathfrak{C}^{ijkl}\mu^3_{ijk} (2\omega) B_l+ \frac{1}{2!2!}\mathfrak{C}^{ijkl}\mu^2_{ij} B^2_{kl}\right]\wedge u \\
\mathcal{V}_{anom} &= -\mathfrak{C}^{jklm}\mu_j\left[\frac{1}{6} B^3_{klm} + \frac{1}{2} B^2_{kl}\wedge\mu_m\omega\right.\\
&\qquad\left.+\frac{2}{3} B_k\wedge \mu^2_{lm}\omega^2+\frac{1}{3}\mu^3_{klm}\omega^3\ \right]\wedge u \\
\bar{q}_{anom} &=\frac{1}{2}\frac{\delta \mathcal{V}_{anom}}{\delta \omega}\wedge u\\
&= -\mathfrak{C}^{ijkl}\mu^2_{ij}\left[\frac{1}{4} B^2_{kl} + \frac{2}{3} B_k\wedge\mu_l\omega+\frac{1}{2}\mu^2_{kl}\omega^2 \right]\wedge u \\
\bar{J}^i_{anom} &= \frac{\delta\mathcal{V}_{anom}}{\delta B_i}\wedge u\\
&=-\mathfrak{C}^{ijkl}\mu_j\left[\frac{1}{2} B^2_{kl} +  B_k\wedge\mu_l\omega+\frac{2}{3}\mu^2_{kl}\omega^2 \right]\wedge u \\
\bar{J}_{S,anom} &= \frac{1}{2\omega}\frac{\partial \mathcal{V}_{anom}}{\partial T}\wedge u=0
\end{split}
\end{equation}
\subsection*{d=8}
\begin{equation*}
\begin{split}
\bar{\mathfrak{A}}^i &= \frac{1}{4!}\mathfrak{C}^{ijklm}F^4_{jklm} \\
\mathcal{V}_{anom} &=  -\mathfrak{C}^{jklmn}\mu_j\left[\frac{1}{24} B^4_{klmn} +\frac{1}{6} B^3_{klm} \wedge\mu_n\omega \right.\\
& \left. +  \frac{1}{3} B^2_{kl}\wedge\mu^2_{mn}\omega^2 +\frac{1}{3} B_k\wedge \mu^3_{lmn}\omega^3 +\frac{2}{15}\mu^4_{klmn}\omega^4 \right]\wedge u \\
\bar{\mathcal{G}}_{anom} &=\left[\frac{1}{5!}\mathfrak{C}^{ijklm}\mu^5_{ijklm} (2\omega)^3+\frac{1}{4!}\mathfrak{C}^{ijklm}\mu^4_{ijkl} (2\omega)^2 B_m\right.\\
&\left. +\frac{1}{3!2!}\mathfrak{C}^{ijklm}\mu^3_{ijk} (2\omega) B^2_{lm}+ \frac{1}{2!3!}\mathfrak{C}^{ijklm}\mu^2_{ij} B^3_{klm}\right]\wedge u \\
\end{split}
\end{equation*}
\begin{equation}\label{eq:soln8}
\begin{split}
\bar{q}_{anom} &=\frac{1}{2}\frac{\delta \mathcal{V}_{anom}}{\delta \omega}\wedge u\\
&= -\mathfrak{C}^{ijklm}\mu^2_{ij}\left[\frac{1}{12} B^3_{klm} + \frac{1}{3} B^2_{kl}\wedge\mu_m\omega\right.\\
&\qquad\left.+\frac{1}{2} B_k\wedge \mu^2_{lm}\omega^2+\frac{4}{15}\mu^3_{klm}\omega^3 \right]\wedge u \\
\bar{J}^i_{anom} &= \frac{\delta\mathcal{V}_{anom}}{\delta B_i}\wedge u \\
&= -\mathfrak{C}^{ijklm}\mu_j\left[\frac{1}{6} B^3_{klm} + \frac{1}{2} B^2_{kl}\wedge\mu_m\omega\right.\\
&\qquad\left.+\frac{2}{3} B_k\wedge \mu^2_{lm}\omega^2+\frac{1}{3}\mu^3_{klm}\omega^3 \right]\wedge u \\
\bar{J}_{S,anom} &= \frac{1}{2\omega}\frac{\partial \mathcal{V}_{anom}}{\partial T}\wedge u=0
\end{split}
\end{equation}
\subsection*{d=10}
\begin{equation*}
\begin{split}
\bar{\mathfrak{A}}^i &= \frac{1}{5!}\mathfrak{C}^{ijklmn}F^5_{jklmn}\\
\mathcal{V}_{anom} &= -\mathfrak{C}^{jklmnp}\mu_j\left[\frac{1}{120} B^5_{klmnp}+\frac{1}{24} B^4_{klmn}\wedge \mu_p\omega +  \frac{1}{9} B^3_{klm}\wedge \mu^2_{np}\omega^2\right.\\
&\qquad\left.+ \frac{1}{6} B^2_{kl} \wedge\mu^3_{mnp}\omega^3 +\frac{2}{15}B_k\mu^4_{lmnp}\omega^4 +\frac{2}{45}\mu^5_{klmnp}\omega^5\right] \\
\bar{\mathcal{G}}_{anom} &=\left[\frac{1}{6!}\mathfrak{C}^{ijklmn}\mu^6_{ijklmn} (2\omega)^4+\frac{1}{5!}\mathfrak{C}^{ijklmn}\mu^5_{ijklm} (2\omega)^3 B_n\right.\\
&\left. +\frac{1}{4!2!}\mathfrak{C}^{ijklmn}\mu^4_{ijklm} (2\omega)^2 B^2_{mn}+\frac{1}{3!3!}\mathfrak{C}^{ijklmn}\mu^3_{ijk} (2\omega) B^3_{lmn}\right.\\
&\left.+ \frac{1}{2!4!}\mathfrak{C}^{ijklmn}\mu^2_{ij} B^4_{klmn}\right]\wedge u \\
\bar{q}_{anom} &=\frac{1}{2}\frac{\delta \mathcal{V}_{anom}}{\delta \omega}\wedge u = -\mathfrak{C}^{ijklmn}\mu_i\mu_j\left[\frac{1}{48} B^4_{klmn}+\frac{1}{9} B^3_{klm} \wedge\mu_n\omega\right.\\
&\qquad \left. +  \frac{1}{4} B^2_{kl}\wedge\mu^2_{mn}\omega^2 +\frac{4}{15} B_k\wedge \mu^3_{lmn}\omega^3 +\frac{1}{9}\mu^4_{klmn}\omega^4 \right]\wedge u \\
\end{split}
\end{equation*}

\begin{equation}\label{eq:soln10}
\begin{split}
\bar{J}^i_{anom} &= \frac{\delta\mathcal{V}_{anom}}{\delta B_i}\wedge u = -\mathfrak{C}^{ijklmn}\mu_j\left[\frac{1}{24} B^4_{klmn}+\frac{1}{6} B^3_{klm}\wedge\mu_n\omega \right.\\
&\qquad\left.+  \frac{1}{3} B^2_{kl}\wedge\mu^2_{mn}\omega^2+\frac{1}{3} B_k\wedge \mu^3_{lmn} \omega^3 +\frac{2}{15} \mu^4_{klmn}\omega^4 \right]\wedge u \\
\bar{J}_{S,anom} &= \frac{1}{2\omega}\frac{\partial \mathcal{V}_{anom}}{\partial T}\wedge u=0
\end{split}
\end{equation}

 \section{Notation}\label{app:notation}
We work in the $(-++\ldots)$ signature. The dimensions of the spacetime in which the fluid lives is denoted by $d=2n$.
The Greek indices $\mu,\nu= 0,1,\ldots,d-1$ are used as space-time indices, whereas the Latin indices $i,j,k\ldots$ are
used as the flavor charge indices. .

We denote Hodge-duals by an overbar - for example, $\bar{J}_i$ is the 2n-1 form Hodge-dual to the
1-form $J_i$. We mostly just use the Hodge-duality between 1-forms and 2n-1 forms and our conventions
are completely defined by  the following statement- given any $2n-1$ form $\bar{V}$ hodge-dual to $V_\mu$
and a 1-form $A_\mu$, we have 
\begin{equation}
\begin{split}
D\bar{V}&=(D_\mu V^\mu)\ \text{Vol}_{2n}\\
A\wedge\bar{V}&=-\bar{V}\wedge A = A_\mu V^\mu \ \text{Vol}_{2n}\\
\end{split}
\end{equation}
Given a 0-form $\alpha$ its Hodge-dual 2n-form is simply $\bar{\alpha}\equiv \alpha \ \text{Vol}_{2n}$.  

We have included a table with other useful parameters used in the text. In the table~\ref{notation:tab}, the relevant equations are denoted by their respective equation numbers appearing inside parentheses.

\begin{table}\label{notation:tab}
 \centering
 \begin{tabular}{||r|l||r|l||}
   \hline
   \multicolumn{4}{||c||}{\textbf{Table of Notation}} \\
   \hline 
   Symbol & Definition & Symbol & Definition \\
   \hline
   $u^\mu,u$ & Fluid velocity, 1-form  &  & \\
   $g_{\mu\nu}$ & Spacetime metric & $P_{\mu\nu}$ & $g_{\mu\nu}+u_\mu u_\nu$ \\
   $\varepsilon$ & Fluid energy density & $p$ & Fluid pressure \\
   $n^i$ & Fluid charge density & $s$ & Fluid entropy density \\
   $\mu_i$& Chemical potentials  & $T$ & Fluid temperature\\
   $T^{\mu\nu}$ & Energy-momentum  & $J^{i\mu}$ & Charge currents\\
    & tensor of the fluid &  &  with anomalies \\
   $J^\mu_S $ & Entropy current & $\mathfrak{C}^{ij\ldots}$ & Anomaly coefficient  \\
   $q^{\mu}_{anom}$ & Anomaly-induced  & $\bar{q}_{anom}$ & Hodge-dual of $q^{\mu}_{anom}$  \\
                    & heat current    &  & $2n-1$ form \\
    $J^{i\mu}_{anom}$ & Anomaly-induced & $\bar{J}^{i}_{anom}$ & Hodge-dual of $J^{i\mu}_{anom}$ \\
        & Charge current &  & $2n-1$ form \\
   $J^\mu_{S,anom} $ & Anomaly-induced  & $\bar{J}_{S,anom}$ &  Hodge-dual of $J^\mu_{S,anom} $\\
                     &Entropy current &  & $2n-1$ form \\
   $F_{i\mu\nu},F_i$ & non-dynamical gauge   & $E_i^\mu,E_i$ & Rest frame electric \\
               & field strength, 2-form &  & field $F_{i\mu\nu}u^\nu$, 1-form \\
   $B_{i\mu\nu},B_i$ & Rest frame magnetic & $\mathfrak{A}^i$ & Anomaly in the  \\
               &  fields $F_i-u\wedge E_i$ &  & ith current $D\bar{J}^i\equiv\mathfrak{A}^i$ \\
   $a_\mu,a$ & Acceleration field & $D_\mu$ & Flavor/Lorentz   \\
   &  $(u.D)u_\mu$, 1-form  & & Covariant derivative\\
   $\sigma_{\mu\nu}$ & Shear strain rate  & $\omega_{\mu\nu},\omega$ & Fluid vorticity, 2-form \\
   $\mathcal{V}_{anom}$ & A formal spatial 2n-form & $\bar{\mathcal{G}}_{anom}$ & Anomaly-induced  \\
   & encoding anomalous transport & & the Gibbs current (Hodge dual)  \\
  \hline
\end{tabular}
\end{table}

\bibliographystyle{JHEP}
\bibliography{cvebibv2}

\providecommand{\href}[2]{#2}\begingroup\raggedright\begin{thebibliography}{10}

\bibitem{Kovtun:2004de}
P.~Kovtun, D.~T. Son, and A.~O. Starinets, {\it {Viscosity in strongly
  interacting quantum field theories from black hole physics}},  {\em Phys.
  Rev. Lett.} {\bf 94} (2005) 111601,
  [\href{http://arxiv.org/abs/hep-th/0405231}{{\tt hep-th/0405231}}].

\bibitem{Bhattacharyya:2008jc}
S.~Bhattacharyya, V.~E. Hubeny, S.~Minwalla, and M.~Rangamani, {\it {Nonlinear
  Fluid Dynamics from Gravity}},  {\em JHEP} {\bf 02} (2008) 045,
  [\href{http://arxiv.org/abs/0712.2456}{{\tt arXiv:0712.2456}}].

\bibitem{Bhattacharyya:2007vs}
S.~Bhattacharyya, S.~Lahiri, R.~Loganayagam, and S.~Minwalla, {\it {Large
  rotating AdS black holes from fluid mechanics}},  {\em JHEP} {\bf 09} (2008)
  054, [\href{http://arxiv.org/abs/0708.1770}{{\tt arXiv:0708.1770}}].

\bibitem{Erdmenger:2008rm}
J.~Erdmenger, M.~Haack, M.~Kaminski, and A.~Yarom, {\it {Fluid dynamics of
  R-charged black holes}},  {\em JHEP} {\bf 01} (2009) 055,
  [\href{http://arxiv.org/abs/0809.2488}{{\tt arXiv:0809.2488}}].

\bibitem{Banerjee:2008th}
N.~Banerjee {\em et~al.}, {\it {Hydrodynamics from charged black branes}},
  {\em JHEP} {\bf 01} (2011) 094, [\href{http://arxiv.org/abs/0809.2596}{{\tt
  arXiv:0809.2596}}].

\bibitem{Torabian:2009qk}
M.~Torabian and H.-U. Yee, {\it {Holographic nonlinear hydrodynamics from
  AdS/CFT with multiple/non-Abelian symmetries}},  {\em JHEP} {\bf 08} (2009)
  020, [\href{http://arxiv.org/abs/0903.4894}{{\tt arXiv:0903.4894}}].

\bibitem{Son:2009tf}
D.~T. Son and P.~Surowka, {\it {Hydrodynamics with Triangle Anomalies}},  {\em
  Phys. Rev. Lett.} {\bf 103} (2009) 191601,
  [\href{http://arxiv.org/abs/0906.5044}{{\tt arXiv:0906.5044}}].

\bibitem{Neiman:2010zi}
Y.~Neiman and Y.~Oz, {\it {Relativistic Hydrodynamics with General Anomalous
  Charges}},  {\em JHEP} {\bf 03} (2011) 023,
  [\href{http://arxiv.org/abs/1011.5107}{{\tt arXiv:1011.5107}}].

\bibitem{Amado:2011zx}
I.~Amado, K.~Landsteiner, and F.~Pena-Benitez, {\it {Anomalous transport
  coefficients from Kubo formulas in Holography}},
  \href{http://arxiv.org/abs/1102.4577}{{\tt arXiv:1102.4577}}.

\bibitem{Landsteiner:2011cp}
K.~Landsteiner, E.~Megias, and F.~Pena-Benitez, {\it {Gravitational Anomaly and
  Transport}},  \href{http://arxiv.org/abs/1103.5006}{{\tt arXiv:1103.5006}}.

\bibitem{Vilenkin:1978hb}
A.~Vilenkin, {\it {Parity violating currents in thermal radiation}},  {\em
  Phys. Lett.} {\bf B80} (1978) 150--152.

\bibitem{Vilenkin:1979ui}
A.~Vilenkin, {\it {Macroscopic parity violating effects: neutrino fluxes from
  rotating black holes and in rotating thermal radiation}},  {\em Phys. Rev.}
  {\bf D20} (1979) 1807--1812.

\bibitem{Vilenkin:1980fu}
A.~Vilenkin, {\it {Equilibrium parity violating current in a magnetic field}},
  {\em Phys. Rev.} {\bf D22} (1980) 3080--3084.

\bibitem{Vilenkin:1980zv}
A.~Vilenkin, {\it {Quantum field theory at finite temperature in a rotating
  system}},  {\em Phys. Rev.} {\bf D21} (1980) 2260--2269.

\bibitem{Vilenkin:1980ft}
A.~Vilenkin, {\it {Cancellation of equilibrium parity violating currents}},
  {\em Phys. Rev.} {\bf D22} (1980) 3067--3079.

\bibitem{Vilenkin:1995um}
A.~Vilenkin, {\it {Parity nonconservation and neutrino transport in magnetic
  fields}},  {\em Astrophys. J.} {\bf 451} (1995) 700--702.

\bibitem{Kharzeev:2011ds}
D.~E. Kharzeev and H.-U. Yee, {\it {Anomalies and time reversal invariance in
  relativistic hydrodynamics: the second order and higher dimensional
  formulations}},  \href{http://arxiv.org/abs/1105.6360}{{\tt
  arXiv:1105.6360}}.

\bibitem{Bardeen:1984pm}
W.~A. Bardeen and B.~Zumino, {\it {Consistent and Covariant Anomalies in Gauge
  and Gravitational Theories}},  {\em Nucl.Phys.} {\bf B244} (1984) 421.
  Revised version.

\bibitem{Chu:1996fr}
C.-S. Chu, P.-M. Ho, and B.~Zumino, {\it {Non-Abelian Anomalies and Effective
  Actions for a Homogeneous Space $G/H$}},  {\em Nucl. Phys.} {\bf B475} (1996)
  484--504, [\href{http://arxiv.org/abs/hep-th/9602093}{{\tt hep-th/9602093}}].

\bibitem{futurePiotr}
R.~Loganayagam and P.~Surowka, {\it {To appear}}, .

\end{thebibliography}\endgroup
\end{document}